\documentclass[runningheads]{llncs}
\usepackage{graphicx}
\usepackage{hyperref}
\usepackage{xcolor}

\usepackage{caption}
\usepackage{subcaption}
\usepackage{changes}
\usepackage{multirow}

\begin{document}
\title{Glioma Classification Using Multimodal Radiology and Histology Data}
\titlerunning{Glioma Classification Using Multimodal Data}

\author{Azam Hamidinekoo\inst{1,2}\and
Tomasz Pieciak \inst{3,4} \and
Maryam Afzali\inst{5} \and
Otar Akanyeti\inst{6} \and
Yinyin Yuan\inst{1,2}
}
\authorrunning{A. Hamidinekoo et al.}

\institute{Division of Molecular Pathology, Institute of Cancer Research (ICR), London, UK \and
Centre for Evolution and Cancer, Institute of Cancer Research (ICR), London, UK \and
LPI, ETSI Telecomunicaci\'{o}n, Universidad de Valladolid, Valladolid, Spain \and
AGH University of Science and Technology, Krak\'{o}w, Poland  \and
CUBRIC, School of Psychology, Cardiff University, Cardiff, UK \and
Department of Computer Science, Aberystwyth University, Ceredigion, UK}
\maketitle              
\begin{abstract}
Gliomas are brain tumours with a high mortality rate. There are various grades and sub-types of this tumour, and the treatment procedure varies accordingly. Clinicians and oncologists diagnose and categorise these tumours based on visual inspection of radiology and histology data. However, this process can be time-consuming and subjective.
The computer-assisted methods can help clinicians to make better and faster decisions. 
In this paper, we propose a pipeline for automatic classification of gliomas into three sub-types: oligodendroglioma, astrocytoma, and glioblastoma, using both radiology and histopathology images. The proposed approach implements distinct classification models for radiographic and histologic modalities and combines them through an ensemble method. The classification algorithm initially carries out tile-level (for histology) and slice-level (for radiology) classification via a deep learning method, then tile/slice-level latent features are combined for a whole-slide and whole-volume sub-type prediction. The classification algorithm was evaluated using the data set provided in the CPM-RadPath 2020 challenge. The proposed pipeline achieved the F1-Score of 0.886, Cohen's Kappa score of 0.811 and Balance accuracy of 0.860. The ability of the proposed model for end-to-end learning of diverse features enables it to give a comparable prediction of glioma tumour sub-types.

\end{abstract}

\section{Introduction}
Gliomas are tumours of the brain parenchyma which are typically graded from I (low severity) to IV (high severity), and the five-year survival rates are $94\%$ (for grade I) and $5\%$ (for higher grades)~\cite{kumar2018robbins}. Magnetic Resonance Imaging (MRI) has been widely used in examining gliomas during diagnosis, surgical planning and follow-up. Clinical protocols include T1-weighted (T1w), T2-weighted (T2w), fluid-attenuated inversion recovery (FLAIR), and gadolinium-enhanced T1-Weighted (Gd-T1w) imaging. T1w, T2w and FLAIR show the tumour lesion and oedema whereas Gd-T1w shows regions of blood-brain-barrier disruption~\cite{abdalla2020diagnostic}. However, due to the complex tumour micro-environment and spatial heterogeneity, MRI itself is not sufficient for complete characterisation of gliomas (e.g. grading and sub-typing) and for high-grade cases  histopathology examination is often required~\cite{dhermain2010advanced}. In histopathology, gliomas are classified based on the morphological features of the glial cells including increased cellularity, vascular proliferation, necrosis, and infiltration into normal brain parenchyma~\cite{kumar2018robbins,ye2020diffusion}.
Oncologists examine patients’ medical history, radiology scans, pathology slides and reports to provide suitable medical care for a person diagnosed with cancer. This decision making is often subjective and time-consuming. Machine learning offers powerful tools (e.g. deep learning methods) to support automated, faster and more objective clinical decision making. One active area of research is to design data processing pipelines that can effectively combine imaging data at different spatial domains (e.g. micro-scale histology and macro-scale MRI data). These pipelines would enable image-based precision medicine to improve treatment quality~\cite{hamidinekoo2018deep,jameson2015precision}. Recent efforts in automatic classification of gliomas using histology and radiology was reviewed by Kurc et al.~\cite{kurc2020segmentation}. 

In this study, we respond to the CPM-RadPath 2020 challenge and propose a data processing pipeline that classifies gliomas into three sub-types: oligodendroglioma, astrocytoma, and glioblastoma. In our approach, we analysed radiology and histopathology data independently using separate densely connected networks. We then combined the outcomes of each network using a probabilistic ensemble method to arrive at the final sub-type prediction. We used our pipeline to analyse data from a cohort of 35 patients and the results have been submitted to the challenge board. 

\section{Related Work}
In this section, we summarise the work submitted by the participants for the previous similar challenge in CPM-RadPath 2019~(\url{https://www.med.upenn.edu/cbica/cpm-rad-path-2019}). 
Ma et al.~\cite{ma2019brain}, as the first ranked group, proposed two convolutional neural networks to predict the grade of gliomas from both radiology and
pathology data: (i) a 2D ResNet-based model for pathology patch based image classification and (ii) a 3D DenseNet-based model for classifying the detected regions (using a detection model) on multi-parametric MRI (mp-MRI) images. To extract the pathology patches ($512\times512$) they used mean and standard deviation with predefined thresholds to consider patches including cells and excluding patches with background contents. To avoid the effect of intensity variations, they  converted the original RGB  pathology images into gray-scale images. Then the labels of all extracted patches were set to the labels of the entire WSI image. On the radiology side, they used a detection model that was trained on the BraTS2018~(\url{https://www.med.upenn.edu/sbia/brats2018.html}) dataset and used the output of this model (detected abnormality) as the input for the 3D DenseNet model to perform the classification on the MRI volumes. Finally, the prediction results from these two modalities were concatenated using the multinomial logistic regression, resulting in F1-score=0.943 for the validation set. Pei et al.~\cite{pei2019brain} only used the mp-MRI data. They implemented two regular 3D convolutional neural networks (CNN) to analyse MRI data: (i) the first CNN was trained on the  BraTS 2019~(\url{https://www.med.upenn.edu/cbica/brats2019/data.html}) dataset to differentiate tumourous and normal tissues; and (ii) the other CNN was trained on the CPM-RadPath 2019 dataset to do tumour classification. They performed z-score normalization to reduce the impact of intensity inhomogeneity and reported the F1-score=0.764 for the validation set.
Chan et al.~\cite{chan2019automatic} used two neural networks, including VGG16 and ResNet50, to process the patches extracted from the whole slide images. For the radiology analysis, they  used the brain-tumour segmentation method (based on the SegNet) that was developed in the BraTS2018 challenge to obtain the core region of the tumour. Using the predicted regions, they used the pyradiomics module to extract 428 radiomic features. They also used the k-means clustering and random forest to only include relevant pathology patches. Finally, they calculated the sum of the probabilities to produce an ensemble prediction. They compared the prediction results with and without MRI features and found out that the classification models based on MRI radiomic features did not perform as accurate as the whole-slide features. Xue et al.~\cite{xue2019brain} used a dual-path residual convolutional neural network to perform classification. For analysing radiology data, they used their custom designed U-Network to predict tumour region mask
from MRI images. This network was trained on data from the BraTS2019 challenge. Then 3D and 2D ResNet18 CNN architectures were used jointly on detected MRI regions and selected pathology patches (extracted with the pixel resolution of $512\times512$), respectively. They reported the accuracy of  84.9\% on the validation set, and concluded that combining the two image sources yielded a better overall accuracy.

Our proposed method does not use any external data and annotation, and does not require the pre-identification of the abnormality on the radiology data. We use the raw data (without changing them to gray scale or without doing the extensive sub-scaling that reduces the data quality) and feed this data to the models for sub-type prediction. In addition, this approach is not computationally expensive and is fast in the inference phase.

\section{Methodology}
\subsection{Dataset}
To train and validate our pipeline, we used the challenge dataset (\url{https://zenodo.org/record/3718894}). The dataset consists of multi-institutional paired mp-MRI scans and digital pathology whole slide images (WSIs) of brain gliomas, obtained from the same patients. For each subject, a diagnostic classification label, confirmed by pathology,  is provided as `A' (Lower grade astrocytoma, IDH-mutant (Grade II or III)), `O' (Oligodendroglioma, IDH-mutant, 1p/19q codeleted (Grade II or III)) and `G' (Glioblastoma and Diffuse astrocytic glioma, IDH-wildtype (Grade IV)). There are three data subsets: training, validation, and testing with 221, 35, and 73 co-registered radiology and histology
subjects, respectively.  Considering the training subset released by the challenge in the training phase, there were 133, 54, and 34 samples provided for the `G', `O' and `A' classes, respectively.
The radiology data consists of four modalities: T1w,  T2w, Gd-T1w, and FLAIR volumes, acquired with different clinical protocols.  The provided data are distributed after their pre-processing, co-registered to the same anatomical template, interpolated to the same resolution ($1 \ \mathrm{mm}^3$), and skull-stripped.
The histopathology data contains one WSI for each patient, captured from Hematoxylin and Eosin (H\&E) stained tissue specimens. The tissue specimens were scanned with different magnifications and the pixel resolutions.

\subsection{Deep learning model: Densely Connected Network (DCN)}\label{ourmodel}
\begin{figure}[!t]
     \centering
     \begin{subfigure}[b]{\textwidth}
         \centering
         \includegraphics[width=0.95\textwidth]{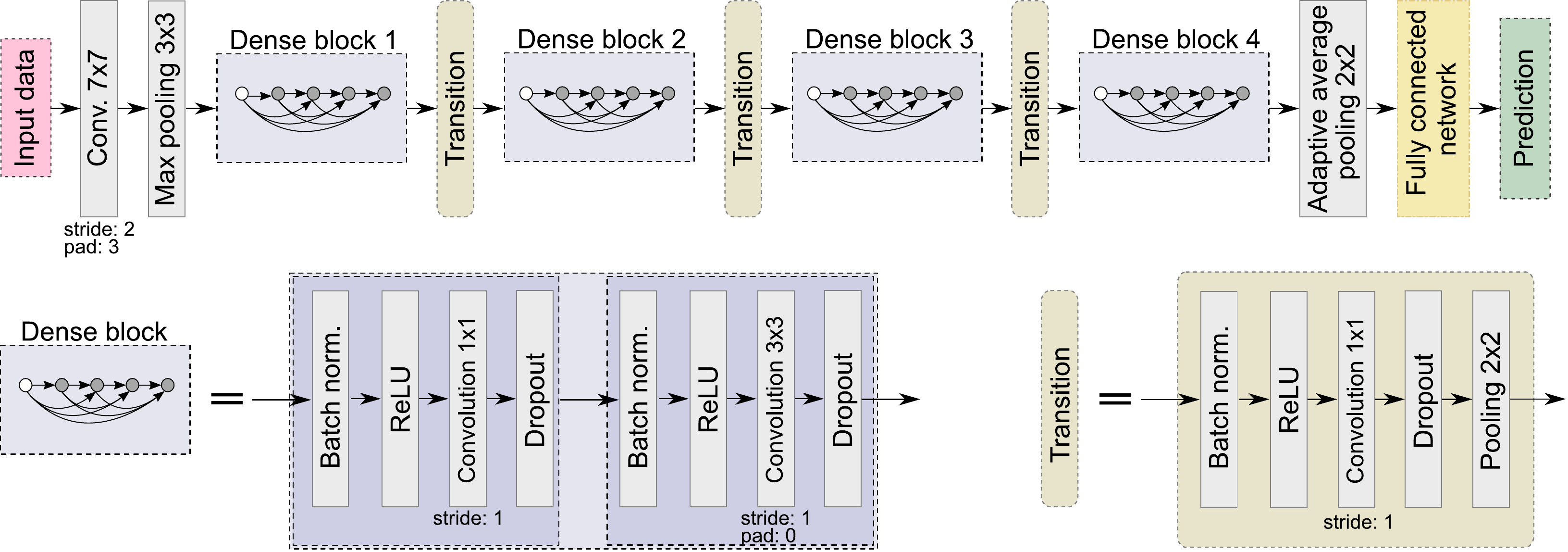}
         \caption{The DCN architecture used for the glioma classification.}
         \label{fig:densenet}
     \end{subfigure}
     \hfill \vspace{0.05cm}
     \begin{subfigure}[b]{\textwidth}
         \centering
         \includegraphics[width=0.95\textwidth]{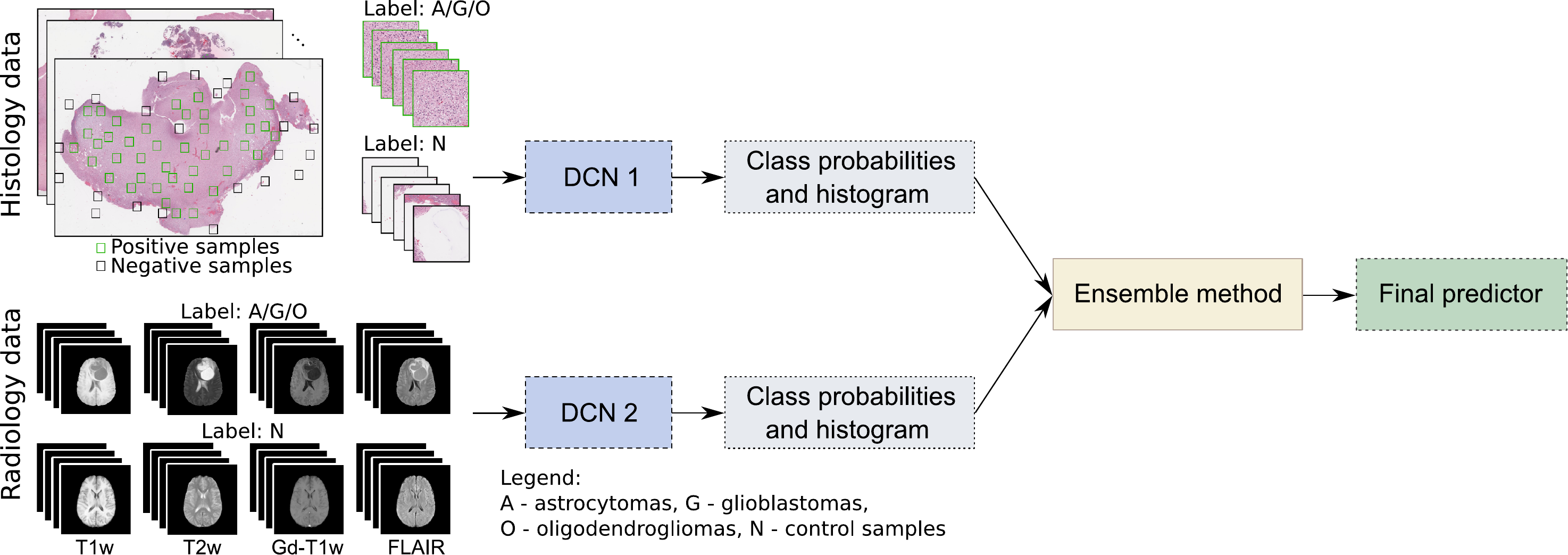}
         \caption{Models trained on different modalities are combined with an ensemble method to predict the patient's glioma sub-types.}
         \label{fig:ensemble}
     \end{subfigure}
        \caption{The proposed pipeline using multi-modality data to classify glioma sub-types.}
        \label{fig:approach}
\end{figure}

Our deep learning classification model is primarily derived from DenseNet architecture~\cite{huang2017densely}. DenseNet is a model notable for its key characteristic of bypassing signals from preceding layers to subsequent layers that enforce optimal information flow in feature maps. The DenseNet model comprises a total of $L$ layers, while each layer is responsible for implementing a specific non-linear transformation. A dense-block consists of multiple densely connected layers and all layers are directly connected. At the end of the last dense-block, a global average pooling is performed to minimize over-fitting and reduce the model's  number of parameters. We have used specific  configurations of this model (called DCN). In the histology DCN (DCN1), the block configuration of [2, 2, 2, 2] was used while in the radiology based DCN (DCN2), the block configuration was set to [6, 12, 36, 24]. The DCN architecture is shown in Fig.~\ref{fig:densenet}. These configurations were selected by training more than 50 models with different block parameters and were empirically chosen. The network growth-rate, which defines the number of filters to add in each layer was set to 32 and 24 for DCN1 and DCN2, respectively. The initial convolution layer contained 64 filters in DCN1 and 48 filters in DCN2 to learn and the multiplicative factor for the number of bottle neck layers was set to 4 with zero dropout rate after each dense layer, as defined in~\cite{huang2017densely}. 

\subsection{Histological phase analysis}\label{pathologyAnalysis}
The process of selecting representative tiles to be used during the training phase is very important and can be labor-intensive. Weakly-supervised approaches~\cite{wang2019weakly} only use the slide labels as supervision during the training of the aggregation model. Some other approaches adopt the multiple instance learning (MIL) method, assuming that the slide label is represented by the existence of positive tiles~\cite{campanella2019clinical}.
To automatically classify  WSIs,  we have adopted a three-stage approach.  In the first stage, 
multiple recognizable tiles/patches of size $2000\times2000$ pixels without any overlaps were selected form the WSI. Each tile was assigned with the slide's label if it included more than 80\% cellularity information. These tiles were counted as positive tiles, which belonged to a specific class of `A', `O' or `G'.
The histology images were representative of slides generated in a true pathology laboratory, including common slide preparation and digitisation artifacts, such as air bubbles, folds, cracks, striping, and blurred regions. To avoid the need for quality control and background removal, we added a new category in the classification problem called `N' representing ``None'' and negative class. All the tiles (i) located on the tissue-background regions (with 80\% background region), (ii) containing any artefacts addressed earlier, and (iii) representing hemorrhage were labeled as the class `N'. So, the tiles with more than 95\% of pixels exceeding 80\% intensity on all three RGB channels were considered. The extracted tiles (positive and negative) were then investigated visually to reassure false tiles were not included in the training set.
Due to the different number of samples provided in each class, we extracted various numbers of positive tiles from WSIs of a class considering the pixel resolution variation. This information is provided in Table~\ref{tab:patchselection}. We selected a balanced number of samples for each class. For example, considering the samples with the 0.25 resolution, we selected 853, 921, 807, and 900 tiles for the `A', `G', `O' and `N' classes, respectively.  In the second stage, we trained the explained DCN on small image tiles extracted from WSIs at high resolution  (see Figure~\ref{fig:ensemble}). Overall, two models were trained: a model for WSIs with pixel resolution of 0.50 and the other model for WSIs with pixel resolution of 0.25. Each tile was then encoded to a feature vector of a low dimension and a prediction score. The output of the DenseNet model was a tile/patch-wise probability prediction. In the third stage, an ensemble approach based on weighted probability prediction major voting was used to integrate the obtained tile level information for whole slide level prediction. For the second stage, we have used the provided slide level labels for the supervision without using any external, extensive, pixel-wise annotations from pathologists. It is important to mention that we assigned the slide label to the extracted patches. This idea of assigning the same label to all extracted patches from the WSI might be criticised since all parts of the WSI might not represent the same information which could lead to a correct classification of the tumour sub-type. To answer this concern, we emphasize that the final prediction was based on the abundance prediction of one of the three desired sub-types.

\begin{table}[b]
        \centering
        \caption{Number of selected samples from each class using the histology data.}
        \label{tab:patchselection}
        \scalebox{0.98}{
        \begin{tabular}{c|cccc}
        \hline
        \multirow{2}{*}{Class}& \multicolumn{2}{c}{Resolution = 0.25} &  \multicolumn{2}{c}{Resolution = 0.50}\\\cline{2-5}
         & Provided cases& {\color{blue}Extracted tiles} & Provided cases & {\color{blue}Extracted tiles}\\  \hline
        O  & 91 & {\color{blue}10} & 42 & {\color{blue}5}\\
        A  & 46 & {\color{blue}10} & 8 & {\color{blue}20}\\
        G  & 31 & {\color{blue}31} & 3 & {\color{blue}50}\\
          \hline
        \end{tabular}}
\end{table}

\subsection{Radiological phase analysis}\label{radiologyAnalysis}
There are several pipelines proposed for automatic prediction of glioma sub-types using MRI data~\cite{kurc2020segmentation}. Most of these approaches are based on analysing radiomics in terms of high-dimensional quantitative features extracted from a large number of medical images. Several approaches initially perform segmentation, followed by the classification based on the detected bounding box of the region of interest~\cite{decuyper2020automated}. The performance of these approaches is affected by the detection and segmentation outcome. We adopted the similar classification scheme used in Section~\ref{pathologyAnalysis} to analyse different MRI modalities. For each modality, the slices in each volume were considered either negative (without any lesions apparent) or positive (with a visible lesion). Each positive slice could belong to one of the target classes (`A', `G', and `O'). With this approach, we avoided the need for lesion segmentation in brain volumes. The initial preparation steps for the MRI slices included (i) visually categorising all the slices provided for each modality into negative and positive and (ii) preparing the training and validation subsets for training with the proportion of 90\% and 10\% of the prepared categorised data, respectively.  Considering that there were fewer samples provided for the `A' class, and the fact that we wanted to include a balanced number of samples in each class during the training, we selected 1500 samples for each class. 
Then we trained a DCN on image slices from each modality volume (see Fig.~\ref{fig:ensemble}). So, four models were trained. Each slice was encoded to the latent feature vector of a low dimension and a~prediction score. The outputs of these models, trained for each modality, were slice-based probability predictions. Subsequently, the weighted average operation was applied to the probability predictions to integrate the obtained slice level information for the whole brain volume. In this classification approach, we have used the provided volume level labels for the supervision without using any external pixel-wise lesion masks from radiologists.

\subsection{Training DCNs}
\begin{figure}[!t]
     \centering
         \centering
         \includegraphics[width=1.0\textwidth]{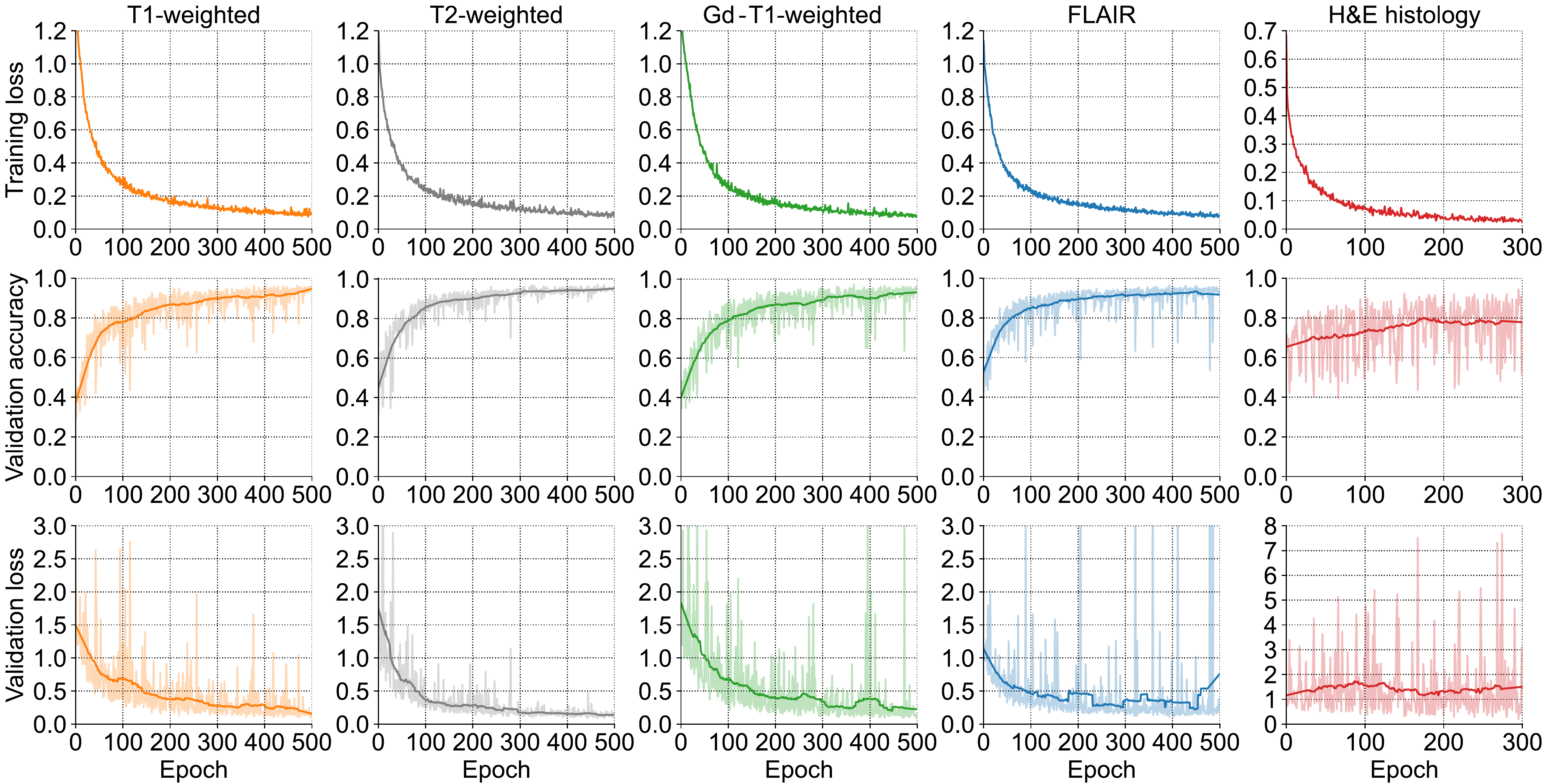}
         \caption{The training process in terms of loss decrease and accuracy improvement for models trained on each modality. Pixel resolution for histology images=0.25.}
         \label{fig:training}
        \label{fig:results}
\end{figure}

We randomly split the provided training dataset (radiology and histology) into training and validation sets (90\% and 10\%, respectively), making sure that the validation samples did not overlap with the training samples. Data augmentation was performed to increase the number of training samples by applying random flipping, rotations, scaling and cropping. The final softmax classifier of the DCN contained 4 output nodes that would predict the probability for each class based on the extracted features in the network. The models were implemented in PyTorch and trained via the Adam optimisation algorithm \cite{kingma2014adam} for the cross-entropy loss with mini-batches of size 128. We trained the histology model (DCN1) for 300 epochs while the radiology based models (DCN2) were trained for 500 epochs.  All models were trained independently using an initial learning rate of 0.001, without any step-down policy and weight-decay. The loss function was calculated for each task on all samples in the batch with known ground truth labels and averaged to a global loss.  Then the predicted loss for the slide label was back-propagated through the model. The training process in terms of loss decrease and accuracy improvement for each of these models is shown in Fig.~\ref{fig:training}. 

\subsection{Outcome integration and final sub-type prediction}
It is common to use the combination of histology and MRI data in deep learning approaches to predict the class of gliomas. In this work, we used a deep learning-based method (DCN) for the classification of gliomas. The dataset in each modality was used for training an independent DCN model. Initially, the outputs of these models were either tile-wise or slice-wise probability predictions. Therefore, an ensemble approach based on probability prediction and major voting was used to integrate the obtained tile-level/slice-level information for the whole histology slide or MRI volume. Then the outcomes from five models (one for histology and four for radiology) were integrated with the same ensemble approach based on confidence prediction and major voting. All modality predictions were used for the final classification prediction of a patient.

\section{Results and Discussion}
We trained five models (4 radiology based models and 1 histology based model when pixel resolution=0.25). The F1-Score of the combined model on the training subset was 0.946. Further evaluation of 
these models were separately performed on the provided validation dataset and the evaluation metrics were calculated by the online web-portal of the challenge. Table~\ref{tab:results} shows the classification performance of these models. These results show that among multiple MRI modalities, T2w was able to extract more salient latent features while hyper-intense FLAIR did not reflect the morphological complexity underlying tumour accurately. The agreement between T1w and Gd-T1w was also observed with regards to the classification performance. However, the model trained on the histology tiles achieved the best performance of F1-Score=0.771 compared to other MRI modalities. This illustrated the existence of an abundant amount of information in the histological images which led to the extraction of more distinct latent features to be used by the softmax classifier, which is crucial for high quality clinical decision support. Combining the outcomes from all the models reduced the F1-Score to 0.714. This was expected because including three modalities (T1w, Gd-T1w, and FLAIR)  with more false positive and false negatives along with the major voting could deviate the prediction agreement and affect the final results. However, using the best predictions from histology, T2w and Gd-T1w achieved the best performance of F1-Score = 0.886.
Apart from the quantified results, we visually investigated several histology samples that were correctly classified in the first validation set (derived from the provided training subset). Based on morphological features, we observed that \textit{astrocytomas} contained hyperchromatic nuclei, with prominent nucleoli. \textit{Oligodendrogliomas} were mostly represented as round nuclei surrounded by a clear cytoplasmic halo (like fried egg-shaped) and \textit{glioblastomas} showed a densely cellular tumour with variation in the gross appearance of the viable tumour along with more necrosis. These observations meet the clinical observations addressed by Bertero et al.~\cite{bertero2019classification}. 
\begin{table}[!t]
    \caption{Classification performance of each model, evaluated on the  validation set provided by the CPM-RadPath 2020 challenge. Pixel resolution for histology images=0.25.}
        \centering
        \renewcommand{\arraystretch}{1.2}        
        \label{tab:results}
        \begin{tabular}{c|c|c|c}
        \hline\hline
        Modality              &	F1-Score & Kappa & Balance accuracy\\ \hline
        MRI FLAIR             &	0.543 & 0.346 & 0.634 \\
        MRI T1w               &	0.400 & 0.143 & 0.429 \\
        MRI Gd-T1w            &	0.600 & 0.433 & 0.723 \\
        MRI T2w               &	0.714 & 0.519 & 0.632 \\
        Histology             &	0.771 & 0.606 & 0.682 \\\hline
        Combined Histology + MRI T2w + MRI Gd-T1w    &	\textbf{0.886} & \textbf{0.811} & \textbf{0.860}\\
        Combined Histology and MRIs &0.714 &  0.554 & 0.723 \\
          \hline \hline
        \end{tabular} 
\end{table}
\section{Conclusion}
In this paper, we proposed using specific versions of DenseNet (called DCN) for sub-typing the glioma into astrocytoma, oligodendroglioma, and glioblastoma classes from the MRI and histology images provided by the CPM-RadPath-2020 organisers. 
The evaluations of the five trained models indicated that the combination of radiographic with histologic image information can improve glioma classification performance when careful training is performed in terms of curated data, model architecture, and ensemble approach.

\section*{Acknowledgements}
Azam Hamidinekoo acknowledges the support by Children’s Cancer and Leukaemia Group (CCLGA201906). Tomasz Pieciak acknowledges the Polish National Agency for Academic Exchange for project PN/BEK/2019/1/00421 under the Bekker programme. Maryam Afzali was supported by a Wellcome Trust Investigator Award (096646/Z/11/Z). Otar Akanyeti acknowledges S\'er Cymru Cofund II Research Fellowship grant. Yinyin Yuan acknowledges funding from Cancer Research UK Career Establishment Award (CRUK C45982/A21808), CRUK Early Detection Program Award (C9203/A28770), CRUK Sarcoma Accelerator (C56167/A29363), CRUK Brain Tumour Award (C25858/A28592), Rosetrees Trust (A2714), Children’s Cancer and Leukaemia Group (CCLGA201906), NIH U54 CA217376, NIH R01 CA185138, CDMRP BreastCancer Research Program Award BC132057, European Commission ITN (H2020-MSCA-ITN-2019), and The Royal Marsden/ICR National Institute of Health Research Biomedical Research Centre. This research was also supported in part by PLGrid Infrastructure. Tomasz Pieciak acknowledges Mr. Maciej Czuchry from CYFRONET Academic Computer Centre (Krak\'{o}w, Poland) for outstanding support.

\bibliographystyle{splncs04}
\bibliography{ms}
\end{document}